\documentclass[10pt,english,aps,amssymb,prl,twocolumn, showpacs]{revtex4-1}
\pdfoutput=1
\usepackage[utf8]{inputenc}
\usepackage{amsmath}
\usepackage{graphicx}
\usepackage{amssymb}
\usepackage{esint}
\usepackage{verbatim}

\makeatletter
\@ifundefined{textcolor}{}
{
 \definecolor{WHITE}{gray}{1}
 \definecolor{RED}{rgb}{1,0,0}
 \definecolor{GREEN}{rgb}{0,1,0}
 \definecolor{BLUE}{rgb}{0,0,1}
 \definecolor{CYAN}{cmyk}{1,0,0,0}
 \definecolor{MAGENTA}{cmyk}{0,1,0,0}
 \definecolor{YELLOW}{cmyk}{0,0,1,0}
 }

\renewcommand{\phi}{\varphi}
\renewcommand{\epsilon}{\varepsilon}

\renewcommand{\vec}[1]{{\bf #1}}

\DeclareMathOperator{\IM}{Im}

\DeclareMathOperator{\TR}{Tr}

\newcommand{\bs}{\boldsymbol}
\newcommand{\mc}{\mathcal}

\usepackage{babel}
\begin{document}
\title {Skyrmion-induced subgap states in $p$-wave superconductors}
\author{Kim Pöyhönen$^1$}
\author{Alex Westström$^1$}
\author{Sergey S. Pershoguba$^{2,3}$}
\author{Teemu Ojanen$^1$}
\author{Alexander V. Balatsky$^{2,3}$}
\affiliation{$^1$Department of Applied Physics (LTL), Aalto University, P.~O.~Box 15100,
FI-00076 AALTO, Finland }
\affiliation{$^2$Institute for Materials Science, Los Alamos National Laboratory, Los Alamos, New Mexico 87545, USA}
\affiliation{$^3$Nordita, Center for Quantum Materials, KTH Royal Institute of Technology,
and Stockholm University, Roslagstullsbacken 23, S-106 91 Stockholm, Sweden}
\date{\today}
%
%
%
%
%
%
\begin{abstract}
In $s$-wave systems, it has been theoretically shown that a ferromagnetic film hosting a skyrmion can induce a bound state embedded in the opposite-spin continuum. In this work, we consider a case of skyrmion-induced state in  a $p$-wave superconductor. We find that the skyrmion induces a bound state that generally resides \emph{within} the spectral gap and is isolated from all other states, in contrast to the case of conventional superconductors. To this end, we derive an approximate expression for the $T$-matrix, through which we calculate the spin-polarized local density of states which is observable in scanning tunneling microscopy measurements. We find the unique spectroscopic features of the skyrmion-induced bound state and discuss how our predictions could be employed as novel experimental probes for $p$-wave superconducting states.
\end{abstract}
\maketitle
\bigskip{}
%
%
%
%
%
%
\textit{Introduction ---\hspace{3mm}}Topology has played a significant role in our understanding of robust features of condensed-matter systems. Topological protection guarantees nontrivial materials properties and enables promising potential applications within electronics and technology. Materials with suitable structure provide an excellent ground to produce low-energy excitations analogous to concepts originating in particle physics. One such example is the magnetic skyrmion, a topological defect in a magnetic field which manifests as a vortex-like spin configuration \cite{bogdanov:1989:1,rossler:2006:1}. The magnetic configuration is then characterized by a topological invariant given by
\begin{equation}\label{eq:qcharge}
Q = \frac{1}{4\pi}\int d^2 r \vec{\hat B}\cdot\left(\partial_x \vec{\hat B} \times \partial_y\vec{\hat B}\right),
\end{equation}
where $\hat{\vec{B}}$ is a unit vector aligned with the local magnetic field. $Q$ takes on integer values and is denoted the \textit{topological charge} of the skyrmion. Configurations with different topological charges are separated from each other by a finite energy barrier, making  skyrmions robust excitations.

In recent years, significant experimental progress has been made within the field \cite{muhlbauer:2009:1, munzer:2010:1, yu:2010:1, yu:2011:1, heinze:2011:1, seki:2012:1, neubauer:2009:1, ritz:2013:1,romming:2015:1}. Notably, skyrmions can easily moved by applying spin currents \cite{kiselev:2011:1,zhang:2015:1}. Employing spin-polarized scanning tunneling microscopy, Romming \textit{et al.} \cite{romming:2013:1} demonstrated a controlled method of creating and destroying individual skyrmions. As a consequence, skyrmions are of special interest from a technological perspective due to their properties being potentially suitable for use in electronics.

Concurrently with the expanded experimental possibilities, there has been a rise in interest towards skyrmion-superconductor heterostructures \cite{pershoguba:2016:1,yang:2016:1,hals:2016:1}, not least because the interplay between skyrmions and superconductivity  is expected to give rise to topological systems. In $s$-wave superconductors, skyrmions with $|Q| = 1$ have been theoretically shown to give rise to Yu-Shiba-Rusinov-like states with long-range wavefunctions \cite{pershoguba:2016:1}. These states are within the spectral gap of the bulk states with parallel spin-polarization, but generally still reside within the bulk with anti-parallel spin-polarization, making them resonance peaks in the density of states. Also, Skyrmions with even charge $Q$ have been argued to host Majorana zero-energy states \cite{yang:2016:1} on two-dimensional (2D) $s$-wave substrates.

\begin{figure}
\includegraphics[width = \linewidth]{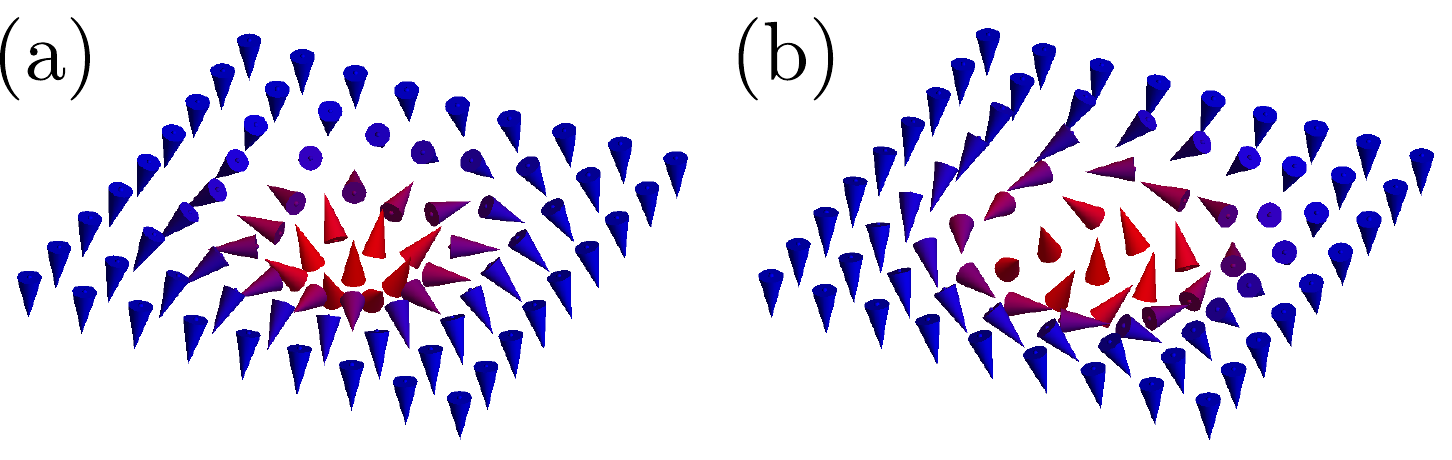}
\caption{Two types of skyrmions with equal topological charge $Q = 1$, a) Néel skyrmion and b) Bloch skyrmion. The respective magnetic field textures are described by Eq.~\eqref{eq:skyrmtexture}.}\label{fig:skyrmions}
\end{figure}

In this letter, we extend the study of skyrmions on superconductors (SCs) to $p$-wave SCs. These are of particular interest due to the rich physics stemming from anisotropic pairing that could itself support topological superconductivity. While this type of unconventional superconductivity has not been conclusively shown to exist in nature, there are nevertheless some possible candidate materials under consideration, most notably Sr$_2$RuO$_4$ \cite{mackenzie:2003:1}. Hence it is interesting to consider what sort of properties such a system would be expected to have. Using standard techniques to analyze the spectra of superconducting states, we find that: \textit{i}) skyrmions with $|Q| = 1$ bind localized subgap states, which is in stark contrast to the results previously obtained for $s$-wave systems; \textit{ii}) depending on the type of $p$-wave coupling, states bound to Bloch and Néel skyrmions show qualitatively different behaviour.

%
%
%
%
%
%
\textit{System -- }The object of interest in this paper is a magnet-SC heterostructure, the magnetic texture being cylindrically symmetric. The simplest case is that of a constant Zeeman field. Consider at first a system consisting of a two-dimensional $p$-wave SC with a constant background magnetic field. The Hamiltonian density of this system is
\begin{equation}
H_0 = \begin{pmatrix}
\xi_p\mathbb{I}_{2\times 2} - B\sigma_z & \Delta \vec{d}\cdot \bs\sigma\\
(\Delta \vec{d}\cdot \bs\sigma)^\dagger & -\xi_p\mathbb{I}_{2\times 2} - B\sigma_z
\end{pmatrix}.
\end{equation}
where $\xi_p = \frac{p^2}{2m} - \mu$ is the kinetic energy, $B$ is the background magnetic field, and $\Delta \vec d\cdot \bs \sigma$ is the superconducting pairing function. This Bogoliubov-de Gennes Hamiltonian acts on the Nambu spinor $\Psi_{\vec k} = (\psi_{\uparrow \vec k},\ \psi_{\downarrow \vec k},\  \psi^\dagger_{\downarrow \vec k},\ - \psi^\dagger_{\uparrow \vec k})^T$, where the operator $\psi^\dagger_{\sigma \vec k}$ creates an electron of spin $\sigma$ and momentum $\vec k$. The superconducting triplet pairing is encoded into the vector $\vec d$. In this paper, we will consider two different types of pairing vectors: out-of-plane $\vec d = (0,\ 0,\ p_x + ip_y)$ and in-plane $\vec d = (p_x,\ p_y,\  0)$. These result in two different Hamiltonians
\begin{equation}\label{eq:hamiltonians}
\begin{split}
H^\text{a}_0 &= \xi_p\tau_z + \Delta(p_x\tau_x-p_y\tau_y)\sigma_z - B\sigma_z\\
H^\text{p}_0 &= \xi_p\tau_z + \Delta(p_x\sigma_x+p_y\sigma_y)\tau_x- B\sigma_z
\end{split}
\end{equation}
where $H^a_0$ is for out-of-plane $\vec d$-vector (pairing of antiparallel spins) and $H^p_0$ for in-plane $\vec d$-vector (pairing of parallel spins); $\tau_i$ and $\sigma_i$ are Pauli matrices in particle-hole and spin space, respectively. Throughout this work we will treat both types of $p$-wave pairing in parallel.

\begin{figure*}
\includegraphics[width=0.32\linewidth]{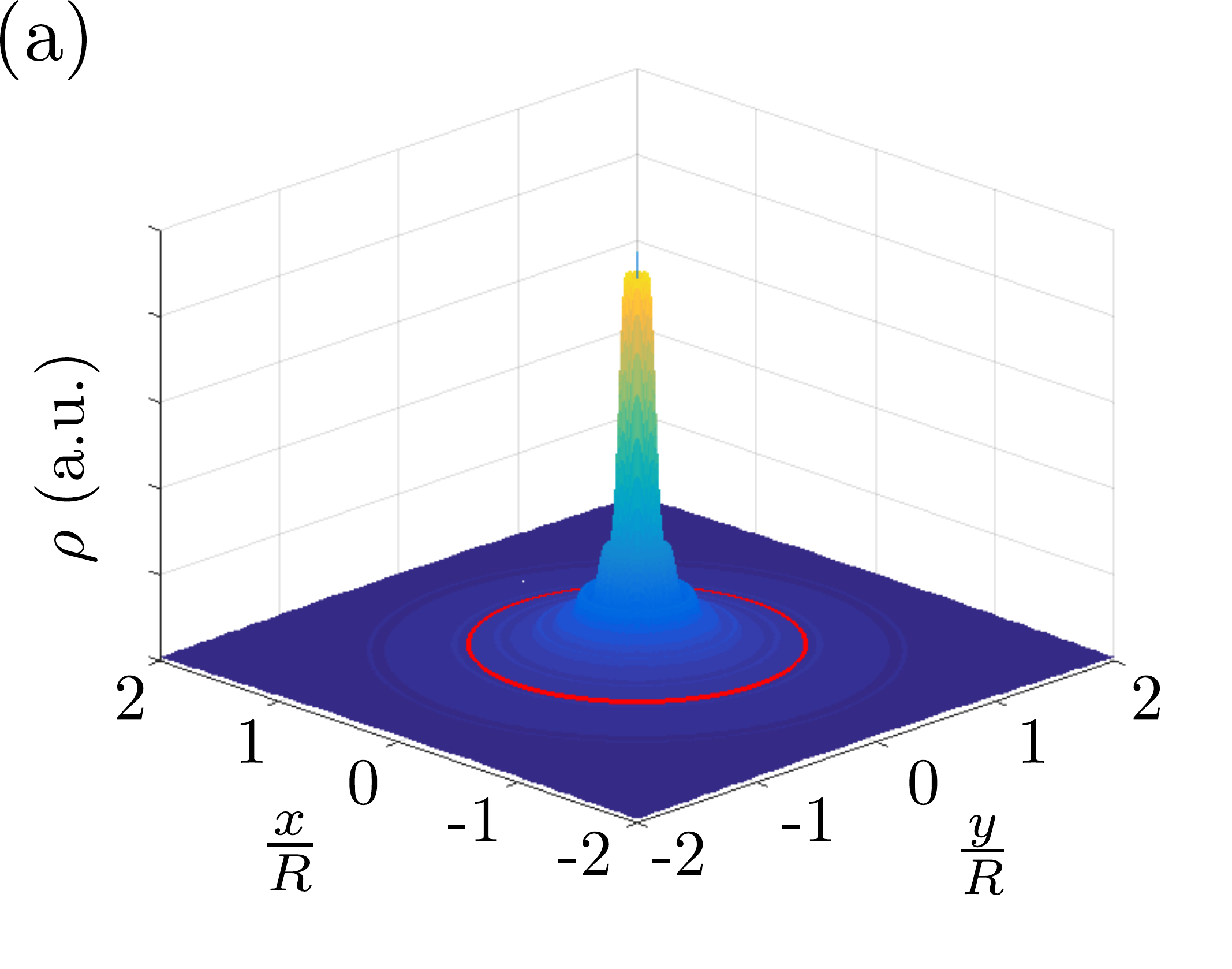}
\includegraphics[width=0.32\linewidth]{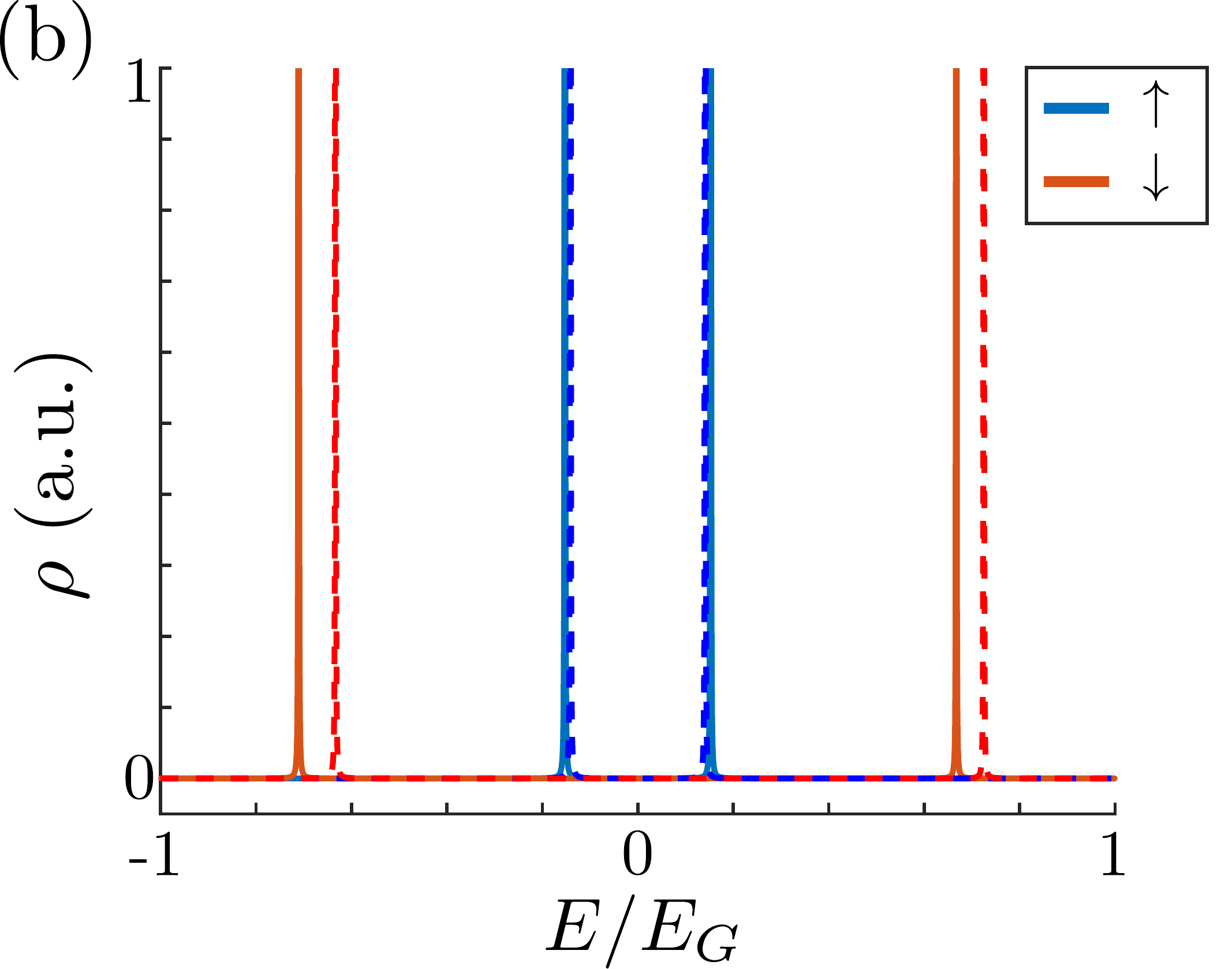}
\includegraphics[width=0.32\linewidth]{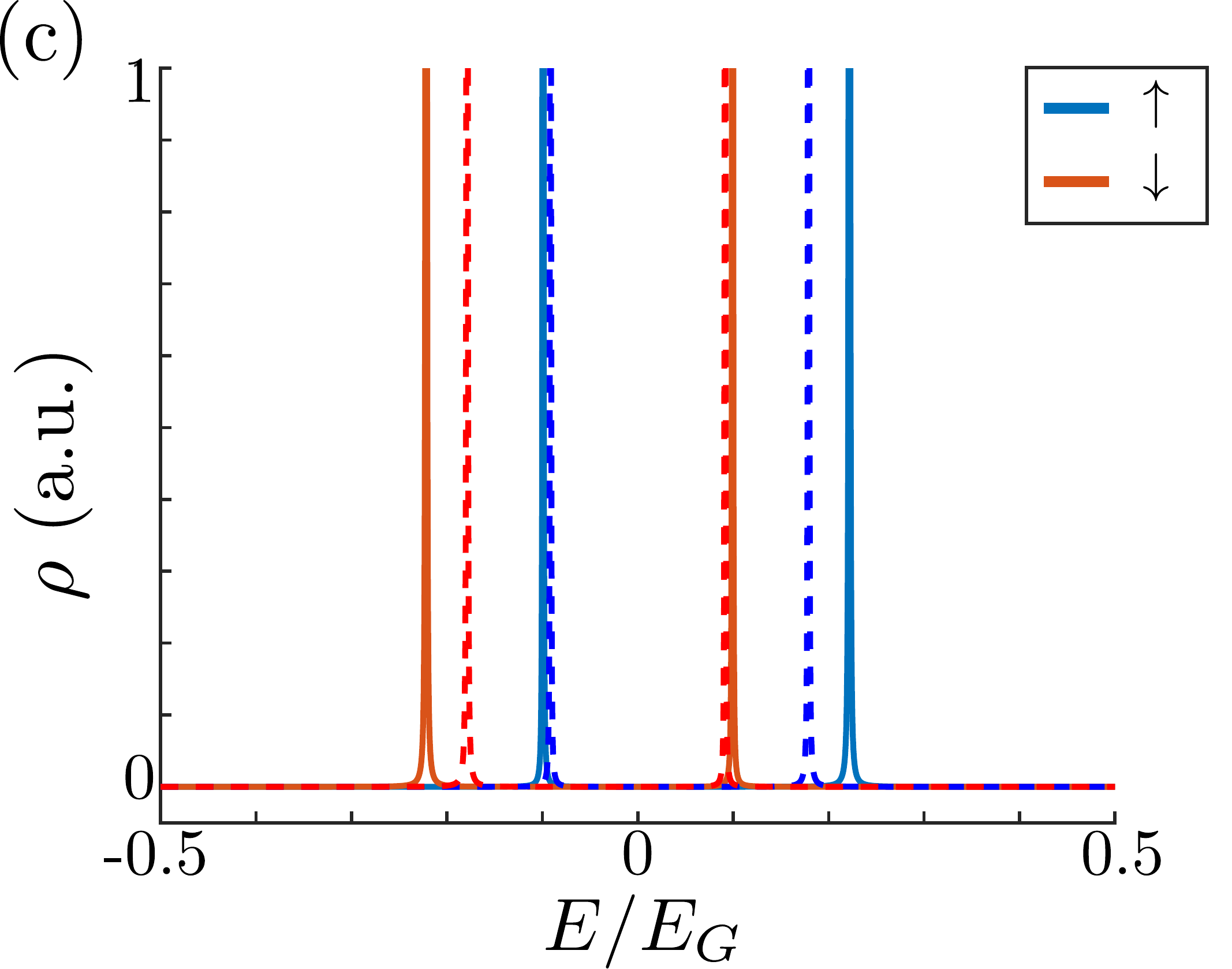}
\caption{a) LDOS of system with a skyrmion and out-of-plane $\vec d$-vector, showing a subgap bound state. Parameters used are $p_F = 20$, $v_F = 100$, $B = \Delta = 0.5$, with skyrmion radius $R = \frac{1}{8}\frac{v_F}{\Delta p_F}$ (red circle). Energy selected is $E = 0.1529E_G$ with an imaginary part of $10^{-3}$ for peak broadening. b) Spin-polarized DOS at $r = 0$ for the same system. The dashed lines correspond to the same parameters but with $U = 4$. Energies are given relative to the bulk gap $E_G$. c) Spin-polarized DOS at $r = 0$ for an in-plane $\vec d$-vector coupled to a Bloch-type skyrmion. Parameters same as in (a) and (b). The dashed line has potential $U = 15$.}\label{fig:LDOS}
\end{figure*}
When the system is coupled to a skyrmion, the magnetic-field term in the Hamiltonian acquires a spatial dependence $B\sigma_z \to \vec B(\vec r)\cdot \bs\sigma$. We will here consider the two different configurations of the magnetic field seen in Fig.~\ref{fig:skyrmions}, known as Néel and Bloch skyrmions, respectively. The fundamental difference between the two cases lies in the direction of rotation for the magnetization vector as a function of radius. The magnetic textures of each is described by the following vectors:
\begin{equation}\label{eq:skyrmtexture}
\begin{split}
\vec B_\text{Néel}(\vec r) &=
B\begin{pmatrix}
\frac{x}{r}\sin\theta(\vec r) & \frac{y}{r}\sin\theta(\vec r) & \cos\theta(\vec r)
\end{pmatrix}\\
\vec B_\text{Bloch}(\vec r) &=
B\begin{pmatrix}
-\frac{y}{r}\sin\theta(\vec r) & \frac{x}{r}\sin\theta(\vec r) & \cos\theta(\vec r)
\end{pmatrix}.
\end{split}
\end{equation}
where we model the position-dependent angle as
\begin{equation}
	\theta(\vec r) =\pi \left\{\begin{array}{c}
		\frac{r}{R},\quad r< R, \\
		1,\quad r\ge R.
	\end{array}\right.
\end{equation}
In the above, $R$ can be viewed as the radius of the skyrmions. Calculating the topological charge for the two configurations above as per Eq.~\eqref{eq:qcharge}, we find that they are equal: for both configurations, $|Q| = 1$. This indicates that the two skyrmions are topologically equivalent, and indeed it is possible to transform between the two through a unitary transformation $H \to e^{-i\pi \sigma_z/4}He^{i\pi \sigma_z/4}$. However, since the transformation between skyrmions does not leave the in-plane $p$-wave Hamiltonian $H^p_0$ invariant, one might expect to see differences between the two types of skyrmions in that model. In order to ascertain the effect of the skyrmions on the $p$-wave system, we will proceed to calculate the local density of states (LDOS), as it contains the relevant spectral properties of the system and further is amenable to experimental analysis.

%
%
%
%
%
%
\textit{Green's function and the multipole expansion ---\hspace{3mm}}
For generic spatial dependence, an explicit analytic solution of the system including the exact magnetic texture is challenging. As a first approximation we  assume that the skyrmion is small compared to the superconducting coherence length. We then perform a multipole expansion around the origin to obtain corrections to a desired order. This gives us an effective potential and allows us to employ the $T$-matrix formalism to find an approximative solution to the full Green's function of the system, the validity of which is tied to that of the multipole expansion. We  restrict ourselves to the second order of the calculation, corresponding to the monopole/anapole term for the Néel/Bloch-type skyrmions, respectively. The expansion for both skyrmions contain a constant magnetic field term, which for the purposes of this work can be treated as part of the unperturbed background; consequently, Eq.~\eqref{eq:hamiltonians} will be used as a starting point of the expansion. The remainder of the multipole terms will then be treated as a scattering potential. We further note that, in real systems, skyrmions can generally be moved around by perturbations unless they are pinned down by static terms \cite{fukuyama:1978:1,lee:1979:1}. This can be modelled by adding a pointlike scalar potential to the skyrmionic terms. Taking this into account, we will consider a change to the Hamiltonian of the form $H \to H_0 - V = H_0 - U\tau_z\delta(\vec r) - \vec B(\vec r)\cdot \bs \sigma$, where
\begin{equation}
\begin{split}
\vec B_\text{Néel}(\vec r) &\approx S_0\sigma_z \delta(\vec r)\vec{\hat z} - S_1\nabla\delta(\vec r)\\
\vec B_\text{Bloch}(\vec r) &\approx S_0\sigma_z\delta(\vec r)\vec{\hat z} - S_1\left(\vec{\hat z}\times\nabla\right)\delta(\vec r)
\end{split}
\end{equation}
Correspondingly, in momentum space, the perturbative potential can be written
\begin{equation}
\begin{split}
V_\text{Néel} &\approx S_0\sigma_z + U\tau_z - iS_1\bs \sigma\cdot\vec p\\
V_\text{Bloch} &\approx S_0\sigma_z + U\tau_z - iS_1\left[\bs \sigma\times \vec p\right]_z.
\end{split}
\end{equation}
The magnitudes of $S_0$ and $S_1$ can be obtained as the respective moments of the expansion. As this calculation makes no reference to the superconductivity, and the skyrmions are generally related by a simple rotation, we can in both cases simply use the magnitudes obtained from the Néel magnetic texture:
\begin{equation}
\begin{split}
S_0 &= \int d^2r [\vec S(\vec r) - \vec S(\infty)]\cdot \vec{\hat z} = \frac{1}{\pi}(\pi^2 - 4)SR^2\\
S_1 &= \frac{1}{2}\int d^2r [\vec S(\vec r) - \vec S(\infty)]\cdot \vec r = \frac{R}{\pi}S_0.
\end{split}
\end{equation}
The above relations fix the values of $S_0$, $S_1$ as a function of the background magnetic field $B$ and the skyrmion radius $R$. Using the truncated multipole expansion we can then find an approximation for the $T$-matrix of the skyrmion through use of the Lippmann-Schwinger equation
\begin{equation}
T(\vec p_1,\vec p_2)\!=\! V(\vec p_1 - \vec p_2) + \int\! \frac{d^2q}{(2\pi)^2} V(\vec p_1 - \vec q)G_0(\vec q) T(\vec q,\vec p_2).
\end{equation}
The equation can be solved analytically in the approximation where the incoming and outgoing momenta are close to the Fermi level \cite{supp}. The explicit solution for the $T$-matrix allows us to calculate the full Green's function of the system.
Inserting our result for the $T$-matrix up to second order in the multipole expansion \cite{supp} results in an expression for the Green's function
\begin{widetext}
\begin{equation}
\begin{split}
G(\vec r) &= G_0(\vec r) + \int \frac{d^2 p^1}{(2\pi)^2} \int \frac{d^2 p^2}{(2\pi)^2} G_0(\vec p^1,\omega)T(\vec{p}^1,\vec{p}^2)G_0(\vec p^2,\omega)e^{i(\vec p^1 -\vec p^2)\cdot \vec r}\\
& \approx G_0(\vec r) + G_0(\vec r)T^0 G_0(-\vec r) + W_i(\vec r) T^1_i G_0(-\vec r) + G_0(\vec r)(T^1_i)^\dagger W_i(-\vec r) + W_i(\vec r)T^2_{ij} W_j(-\vec r),
\end{split}
\end{equation}
\end{widetext}
where we used the fact that the integrals over the two momenta can in each case be separated into two different integrals. Hence finding $G(\vec r)$ reduces to finding the value of the integrals
\begin{align}\label{eq:Wintegral}
G_0(\vec r) & = \int \frac{d\vec p}{2\pi}G_0(\vec p,\omega) e^{i\vec p \cdot \vec r}\\
W_j(\vec r) &= \int \frac{d\vec p}{2\pi}G_0(\vec p,\omega) \frac{p_j}{p}e^{i\vec p \cdot \vec r},
\end{align}
where $G_0(\vec r)$ is the spatial Green's function of the system without a skyrmion. The integrals are analytically tractable, and the solutions for both types of $p$-wave pairing are presented in the supplemental material. We hence have an analytic expression for the full Green's function in terms of these integrals. Moving on, we can use $G(\vec r)$ to calculate the spin-polarized local density of states (SPLDOS) through use of the formula
\begin{equation}\label{eq:SPLDOS}
\rho_\lambda(\vec r) = -\frac{1}{\pi}\IM\TR\left[\frac{1+\lambda\sigma_s}{2}\frac{1+\tau_z}{2} G(\vec r)\right],
\end{equation}
where $\lambda = +1\, (-1)$ corresponds to spin up (down) electrons. The SPLDOS is useful in that it can be probed by spin-polarized STM and hence provides a direct way of comparing theory to experiment. Further, the sum of the terms for spin up and down directly yields the LDOS measured in typical STM experiments.

%
%
%
%
%
%
\begin{figure*}
\includegraphics[width=0.38\linewidth]{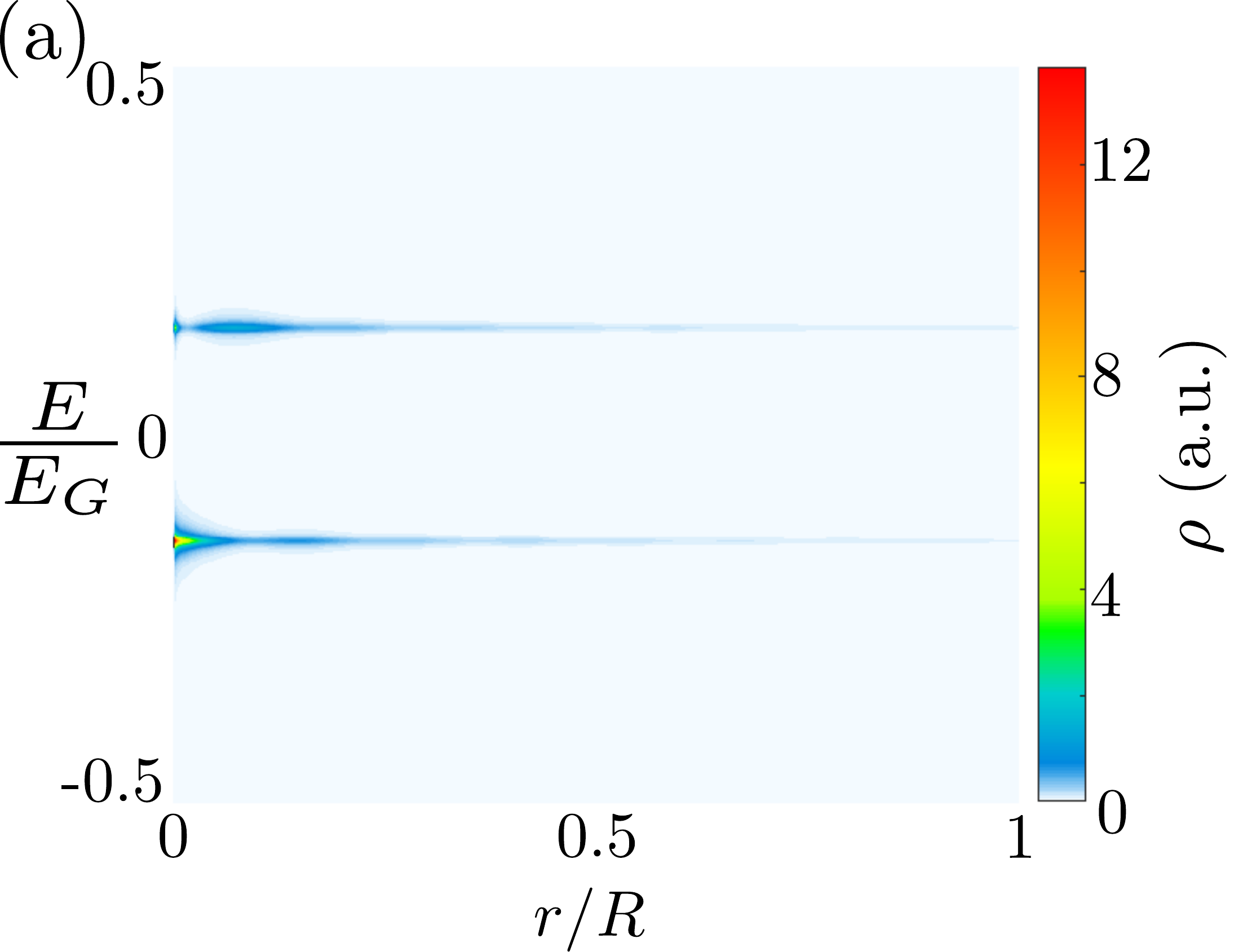}
\hspace{10mm}
\includegraphics[width=0.38\linewidth]{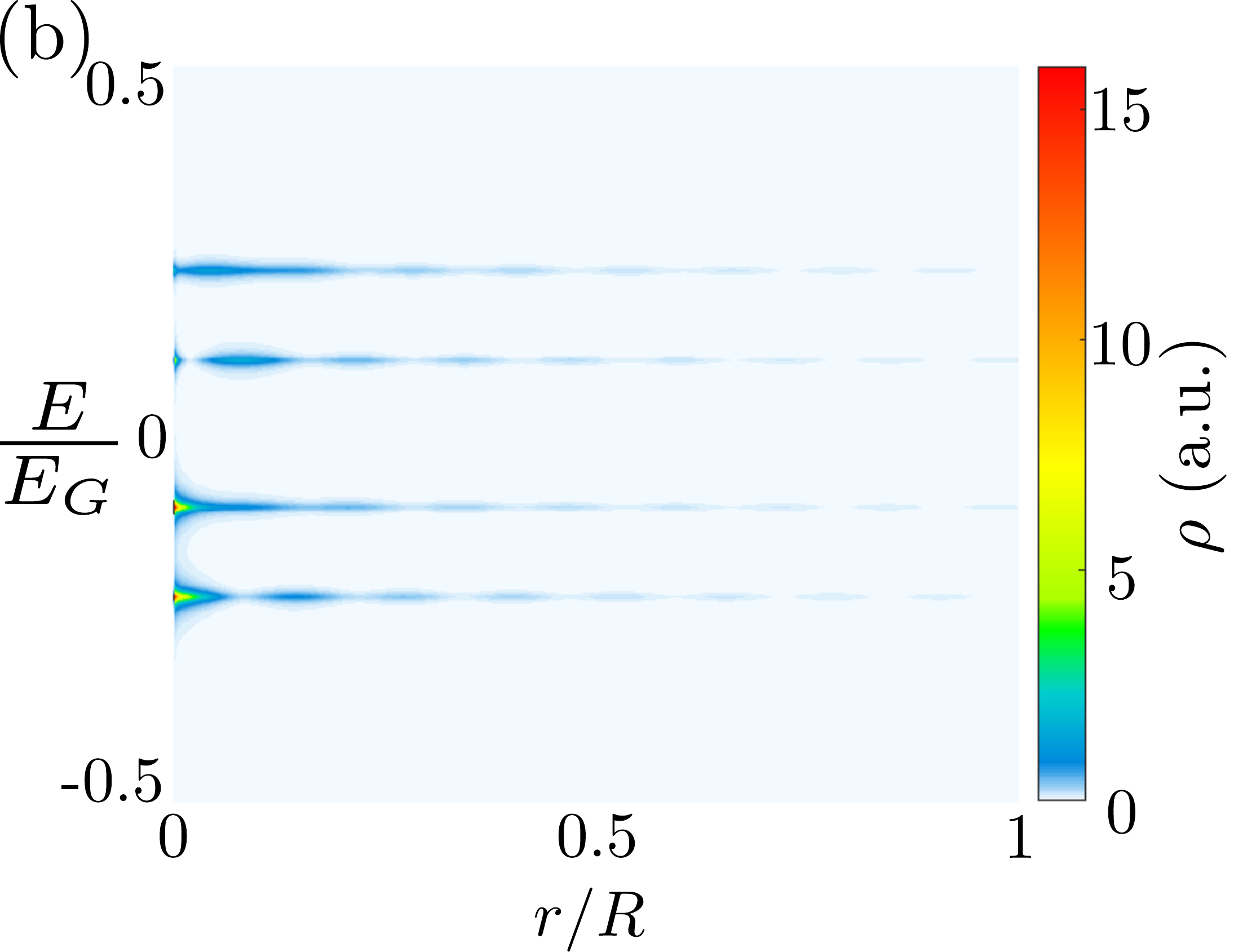}
\caption{LDOS as a function of radius and energy for (a) Out-of-plane $\vec d$-vector; (b) In-plane $\vec d$-vector with a Bloch skyrmion. In both cases, the parameters are $p_F = 20$, $v_F = 100$, $\Delta = B = 0.5$, $R = \frac{1}{8}\frac{v_F}{\Delta p_F}$, and the scalar potential $U = 0$. Energies are given relative to the bulk gap $E_G$.}\label{fig:radialLDOS}
\end{figure*}
\textit{Results ---\hspace{3mm}}~Based on the treatment above, we have calculated the LDOS of the system. Notably, we find that,  the $p$-wave SC-skyrmion system can support subgap bound states well separated from the continuum, unlike the $s$-wave SC, as seen in Fig.~\ref{fig:LDOS}. An example of the LDOS of a system with parameters supporting subgap bound states can be seen in Fig.~\ref{fig:LDOS} (a) for the case of an out-of-plane $\vec d$-vector. To illustrate its location within the gap, in Figs.~\ref{fig:LDOS} (b) and (c) we have plotted the spin-polarized DOS at the origin for both types of $\vec d$-vector. In the latter two we have also included the DOS in the case of non-zero scalar potential $U$ as dashed lines. As is clear from the figures, the  non magnetic potential has a quantitative effect on the energies of pre-existing bound states, although the degree and direction of this shift depends on the $\vec d$-vector. We find in total  four bound states (including those at negative energies), consistent with results obtained for $p$-wave Yu-Shiba-Rusinov systems \cite{kaladzhyan:2016:2}.

Interestingly, the type of skyrmion makes a significant qualitative difference in the case of an in-plane $\vec d$-vector. In the studied regime, the Néel-type skyrmion did not support subgap bound states at all, whereas the Bloch-type skyrmion supports subgap bound states for a wide parameter range \cite{note:skyrmiontrans}. In a system with out-of-plane $\vec d$-vector on the other hand, the two skyrmions are equivalent and the system can support subgap states regardless of skyrmion type. When subgap states are present, they are generally localized within the skyrmion, i.e., $r < R$; the spatial decay of the LDOS is exponential, as illustrated in Fig.~\ref{fig:radialLDOS}, where we have plotted the LDOS as a function of radius and energy for the relevant SC-skyrmion combinations. For completeness, we note that while the in-plane $p$-wave SC coupled to a Néel skyrmion does not support bound states for the parameters used in Figs.~\ref{fig:LDOS} and \ref{fig:radialLDOS}, it can host bound states for high enough scalar potentials $U$. However, in this case the states are clearly bound to the scalar potential rather than the skyrmion, and in fact the presence of the skyrmion increases the scalar potential needed to form a subgap bound state in this system. It is important to note that these features are rather generic: the SC-skyrmion combinations that support subgap bound states do so for a wide range of parameters.

Thus we propose that  it may be possible to distinguish between $s$-wave and different types of $p$-wave superconductivity depending on the effect seen when the system is coupled to skyrmions. The existence of subgap bound states in general indicates that the superconductive pairing is not $s$-wave, and dependence on the type of skyrmion can act as a separator between out-of-plane and in-plane $\vec d$-vectors.

%
%
%
%
%
%
\textit{Conclusions and outlook ---\hspace{3mm}}We have found new bound states that are generated in a $p$-wave
superconductor in proximity to a ferromagnetic film hosting a skyrmion with topological charge $|Q| = 1$. We predict sharp features in the non-polarized as well as  spin-polarized local density of states that can
be measured by tunneling spectroscopy in superconductors. To be general, we considered two types of $p$-wave superconductors, i.e., in-plane and
out-of-plane $\vec d$-vector. In contrast to $s$-wave superconductors, which can only host bound states embedded in a continuum, we found that both studied types of $p$-wave systems can support genuine subgap bound states. In the out-of-plane case, both skyrmions are equivalent due to rotational symmetry. However, for the in-plane $\vec d$-vector, we found remarkable
qualitative differences between the two types of skyrmions: namely, a Bloch skyrmion can induce a subgap bound state in the superconductor for a wide range of parameters, while we observed no bound states for the Neel skyrmion \cite{note:skyrmiontrans}. This feature could be used experimentally to investigate the
character of the pairing of a given $p$-wave superconductor.

We also found that a scalar potential $U$ can have an impact on the structure of the bound states. In general, $U$ will shift the energies of any bound states present, depending on its sign potentially bringing them to low energies or gapping the system altogether. High scalar potentials can result in bound states even for an in-plane $\vec d$ system with a Néel-type skyrmion; however, in this case it is clear that the state is bound specifically to the scalar potential well -- not the skyrmion -- and in fact the presence of the skyrmion increases the minimum scalar potential for which a bound state appears.

Our prediction of subgap bound states also opens up further venues for research, in addition to being useful for distinguishing different types of superconductivity. Within the past decade, it has been experimentally shown that under some circumstances, lattices of skyrmions can form spontaneously \cite{muhlbauer:2009:1, munzer:2010:1, heinze:2011:1}. More recently, 2D lattices of hybridized subgap bound states have been found to give rise to interesting topological behavior in both $s$-wave \cite{rontynen:2015:1, rontynen:2016:1} and $p$-wave \cite{kaladzhyan:2016:1} superconductors. Similarly, we may expect the slowly decaying subgap states in a skyrmion lattice to hybridize and potentially induce interesting topological behavior.

%
%
%
%
%
%
\textit{Acknowledgements ---\hspace{3mm}}
This work was supported by the Academy of Finland and the Aalto Centre for Quantum Engineering (K.P., A.W. and T.O), ITS at ETH Zurich and by US DOE BES E3B7, ERC DM-321031 (S.S.P. and A.V.B.).
%
%
%
%
%
%
\bibliography{SkyrmionBibliography}

\begin{thebibliography}{10}
\bibitem{note:skyrmiontrans}
Note that the transformation between the two skyrmion types changes the
  structure of the in-plane {SC} from the scalar product term used here to a
  vector product-type term, i.e., $\vec p \cdot \boldsymbol\sigma \to [\vec p
  \times \boldsymbol\sigma]_z$. {H}ence in the latter type of system, the
  {N}éel skyrmion contains bound states while the {B}loch skyrmion does not.
\end{thebibliography}
\bibliographystyle{h-physrev}
\begin{widetext}
\appendix
\numberwithin{equation}{section}
%
%
%
%
%
%
\section{$T$-matrix}\label{appendix:t-matrix}
This appendix is dedicated to the derivation of the $T$-matrix for the multipole expansion of the Skyrmion. The derivation done here closely follows that of what was done in the appendix of Ref.~\cite{pershoguba:2016:1}. To simplify, we leave out the scalar impurity $V(\vec k) = -U\tau_z$ throughout the derivation and only reintroduce it at the very end.

Our starting point is the Lippmann-Schwinger equation for the $T$-matrix which reads
\begin{equation}
\begin{split}
T(\vec p^\text{out},\vec p^\text{in}) = &V(\vec p^\text{out} - \vec p^\text{in}) +\\
\int \frac{d^2p}{(2\pi)^2} &V(\vec p^\text{out} - \vec p)G_0(\vec p,\omega) T(\vec p,\vec p^\text{in}).
\end{split}\label{app:eq:LSE}
\end{equation}
In the multipole expansion, the potential is $V(\vec p) = S_0\sigma_z - iS_1f_{ij}p_i\sigma_j$, where the matrix $f_{ij}$ is either equal to the Kronecker delta $\delta_{ij}$ or the Levi-Civita symbol $\epsilon_{3ij}$, depending on whether the Skyrmion is of Néel or Bloch type.

For out-of-plane and in-plane $\vec d$-vector respectively, and as elaborated in Appendix \ref{appendix:integrals}, this is
\begin{equation}\label{eq:G0}
\begin{split}
G_0^a(0,\omega) &= -\pi\nu_0\sum_{\lambda = \pm 1} \mc P^\sigma_\lambda \frac{\omega_\lambda  -\Delta^2\frac{p_F}{v_F\gamma} \tau_z}{\sqrt{\Delta^2 p_F^2-\gamma\omega_\lambda^2}}\\
G_0^p(0,\omega) &= -\pi\nu_0\sum_{\lambda,\lambda^\prime = \pm 1}\mc P^\sigma_\lambda\mc P^\tau_{\lambda^\prime} \frac{\omega - \frac{\Delta^2}{\gamma v_F}(\lambda^\prime p_F + \lambda \frac{B}{v_F})}{\sqrt{\Delta^2\left(\lambda^\prime p_F + \lambda\frac{B}{v_F}\right)^2  - \gamma\omega^2}}
\end{split}
\end{equation}
where $\mc P$ are the projection operators along the $z$ axis -- $\mc P^\sigma_\lambda \equiv \tfrac{1}{2}(1 + \lambda\sigma_z)$, $\mc P^\tau_\lambda \equiv \tfrac{1}{2}(1 + \lambda\tau_z)$ -- and $\gamma \equiv 1 + \Delta^2/v_F^2$. Henceforth we will denote $G_0(0,\omega) \equiv G_0$ for notational simplicity. We can insert $G_0$ into the Lippmann-Schwinger equation in order to calculate the full Green's function of the skyrmion-superconductor composite system:

In order to proceed, we also make a few simplifying observations: first, the potential only consists of a momentum-independent term and a linear term. This observation together with the form of Eq.~\eqref{app:eq:LSE} suggests that a good ansatz for the $T$-matrix is one with terms that are at most quadratic in momentum. The $T$-matrix can then be written as
\begin{equation}
T(\vec p^\text{out},\vec p^\text{in}) = T^0 + T^1_i p^\text{out}_i + (T^1_i)^\dag p^\text{in}_i + T^2_{ij}p^\text{out}_ip^\text{in}_j,\label{app:eq:ansatz}
\end{equation}
where $T^j$ are matrices to be determined. Note that the $(T^1_i)^\dag$ follows from general symmetry arguments for the $T$-matrix. The second observation we make is that the scattering processes primarily occur close to the Fermi surface, allowing us to write $\vec p = p_F \hat{\vec n}$ ($\hat{\vec n}$ is a unit vector and $p_F$ is the Fermi momentum) for all momenta in the ansatz expression for the $T$-matrix and the potential term in Eq.~\eqref{app:eq:ansatz}. Inserting all this into Eq.~\eqref{app:eq:LSE} gives us
\begin{equation}
\begin{split}
&T^0 + T^1_i n^\text{out}_i + (T^1_i)^\dag n^\text{in}_i + T^2_{ij} n^\text{out}_in^\text{in}_j =\\
&V\left(p_F(\vec n^\text{out} - \vec n^\text{in})\right) +  \int \frac{d^2p}{(2\pi)^2} \left[S_0\sigma_z - iS_1p_F f_{ij}(n^\text{out}_i - n_i)\sigma_j\right]G_0(\vec p,\omega)\left[T^0 + T^1_i n_i + (T^1_i)^\dag n^\text{in}_i + T^2_{ij} n_i n^\text{in}_j\right],
\end{split}
\end{equation}
from which we by matching components get a system of equations for the matrix components:
\begin{align}
T^0 &= S_0\sigma_z + S_0\sigma_zG_0(\omega)T^0 + S_0\sigma_z I^1_j T^1_j  I_jT^0\notag\\
&\quad + iS_1p_F f_{jk}\sigma_k + iS_1p_Ff_{ij}\sigma_jI^2_{ik}T^1_k\label{app:eq:T0}\\
T^1_j &= -iS_1p_Ff_{jk}\sigma_k - iS_1p_Ff_{jk}\sigma_k G_0(\omega)T^0\notag\\
&\quad- iS_1p_Ff_{jk}\sigma_k I^1_iT^1_i\label{app:eq:T1}\\
T^2_{ij} &= -iS_1p_Ff_{ik}\sigma_k G_0(\omega) (T^1)^\dagger_j - iS_1 p_Ff_{ik}\sigma_k I^1_lT^2_{lj},\label{app:eq:T2}
\end{align}
where we have introduced the the two integrals
\begin{equation}\label{app:eq:Iintegral}
I^1_j = \int \frac{d^2p}{(2\pi)^2}G_0(\vec p,\omega) n_j,\quad I^2_{ij} = \int \frac{d^2p}{(2\pi)^2}n_iG_0(\vec p,\omega) n_j.
\end{equation}
We postpone the evaluation of these integrals to the next section. The matrix components can now be obtained by first solving for $I^1_iT^1_i$ in Eq.~\eqref{app:eq:T1}. This is achieved by multiplying Eq.~\eqref{app:eq:T1} with $I^1_j$ and summing over $j$. We can then easily express $I^1_iT^1_i$ in terms of $T^0$ and solve the original equation for $T^1_j$ in terms of $T^0$. Inserting this into Eq.~\eqref{app:eq:T0} is then trivial albeit cumbersome. To solve for $T_{ij}^2$ it is then only a matter of performing a similar multiplication and summation trick as we did for $I^1_iT^1_i$. We are finally left with
\begin{align}
T^0 &= \left[Q^\dagger - \Xi G_0(\omega)\right]^{-1}\Xi\\
T^1_j &= -iF_jQ^{-1}(1+G_0(\omega)T^0)\\
T^2_{jk} &= -iF_jQ^{-1}G_0(\omega)(T^1)^\dagger_k,
\end{align}
where we have introduced the matrices
\begin{equation}
\begin{split}\label{app:eq:FQXI}
F_j &= S_1p_Ff_{jk}\sigma_k\\
Q &= 1 + iI^1_jF_j\\
\Xi &= \left[S_0\sigma_z + \frac{1}{2}F_jG_0(\omega)F_j\right]Q^{-1}.
\end{split}
\end{equation}
This concludes the derivation of the $T$-matrix in the multipole expansion. We can now add the scalar impurity term by noting that it would only appear together with $S_0\sigma_z$, so all we need to do is replace $S_0\sigma_z \to S_0\sigma_z + U\tau_z$ in the expression for $\Xi$ in \eqref{app:eq:FQXI} and we are done. As a final note, we emphasize that the derivation is valid for both in- and off-plane $\vec d$, since they only differ in $G_0(\vec p, \omega)$ whose explicit form was not used in the above derivation. 
%
%
%
%
%
%
\section{Integrals}\label{appendix:integrals}
In this Appendix we will evaluate some integrals encountered in the main text and in the previous appendix. Specifically, we consider the integrals in Eq.~\eqref{eq:Wintegral} in the main text,
\begin{align}
G_0(\vec r) &= \int \frac{d\vec p}{2\pi}G_0(\vec p,\omega) e^{i\vec p \cdot \vec r}\\
W_j(\vec r) &= \int \frac{d\vec p}{2\pi}G_0(\vec p,\omega) \frac{p_j}{p}e^{i\vec p \cdot \vec r},
\end{align}
as well as two integrals from the previous appendix:
\begin{align}
I^1_j &= \int \frac{d^2p}{(2\pi)^2}G_0(\vec p,\omega) n_j\\
 I^2_{ij} &= \int \frac{d^2p}{(2\pi)^2}n_iG_0(\vec p,\omega) n_j.
\end{align}
We begin with the integrals from the main text. Due to convergence, it is necessary to consider the cases $r > 0$ and $r = 0$ separately. For the bare Green's function, inverting $H$ and inserting suitable resolutions of identity yields the following integrals at $r = 0$:
\begin{equation}
\begin{split}
G_0^\text{a}(0,\omega) &=\sum_{\lambda = \pm 1}\mc P^\sigma_\lambda \int \frac{d^2 p  }{(2\pi)^2} \frac{\omega + \lambda B + \xi_p\tau_z}{(\omega + \lambda B)^2 - \xi_p^2 -\Delta^2p^2}\\
G_0^\text{p}(0,\omega) &= \sum_{\lambda,\lambda^\prime = \pm 1}\mc P^\sigma_\lambda\mc P^\tau_{\lambda^\prime}\int \frac{d^2 p}{(2\pi)^2} \frac{\omega - \lambda B + \lambda^\prime\xi_p}{\omega^2 - \xi_p^2 - \Delta^2p^2 - B^2 + 2\lambda\lambda^\prime\xi_p B}
\end{split}
\end{equation}
These integrals do not strictly converge as written, ultimately due to the fact that BCS theory is not valid at high energies. They can be calculated, however, by introducing a suitable cutoff. Here, this can simply be effected by assuming the relevant scale is near the Fermi level and hence approximating $p \approx p_F + \xi_p/v_F$. This reduces both Green's functions to simple residue calculations, immediately yielding the expressions in Eq.~\eqref{eq:G0}.

The Green's function for $r > 0$ can be calculated from
\begin{equation}
\begin{split}
G_0^\text{a}(\vec r,\omega) = \sum_{\lambda = \pm 1} \mc P^\sigma_\lambda\int \frac{d^2 p}{(2\pi)^2}  \frac{\omega_\lambda + \xi_p\tau_z + \lambda \Delta (p_x\tau_x - p_y\tau_y)}{\omega_\lambda^2 - \xi_p^2 -\Delta^2p^2}e^{i\vec p \cdot \vec r}\\
G_0^\text{p}(\vec r, \omega) = \sum_{\lambda,\lambda^\prime = \pm 1}\mc P^\sigma_\lambda\mc P^\tau_{\lambda^\prime}\int \frac{d^2 p}{(2\pi)^2}  \frac{\omega + \lambda^\prime\xi_p - \lambda B + \tau_x\Delta( p_x\sigma_x + p_y\sigma_y)}{\omega^2 - \xi_p^2 - \Delta^2p^2 - B^2 + 2\lambda\lambda^\prime\xi_p B}e^{i\vec p \cdot \vec r}.
\end{split}
\end{equation}
The terms in the numerator proportional to the momentum can be written as derivatives with respect to the appropriate coordinate. The angular integral then directly gives a Bessel function of the first kind. Factoring the denominator in terms of its zeroes, we find that calculation of the two Green's functions reduces to solving three integrals:
\begin{align}
G_0^\text{a}(\vec r,\omega) &= \frac{1}{2\pi}\sum_{\lambda = \pm 1} \frac{1 + \lambda\sigma_z}{2} \left[\omega_\lambda I_G^0 + \tau_z I_G^1 + i\lambda \aleph I_G^2\right]\\
G_0^\text{p}(\vec r,\omega) &= \frac{1}{2\pi}\sum_{\lambda,\lambda^\prime = \pm 1}\mc P^\sigma_\lambda\mc P^\tau_{\lambda^\prime}\left[\omega_{-\lambda} I_G^0 + \lambda^\prime I_G^1 +i\Delta\tau_x \bs\sigma\cdot\vec{\hat r} I_G^2 \right]
\end{align}
where we have defined the integrals
\begin{align}\label{app:eq:IG0}
I_G^0 &= \int_0^\infty dp \frac{pJ_0(pr)}{\left[p^2 - (p_0^+)^2\right]\left[p^2-(p_o^-)^2\right]} = -\frac{i}{2\beta}\left[K_0(-ip_0^+r) - K_0(ip_0^-r)\right]\\
I_G^1 &= \int_0^\infty dp \frac{p\xi_pJ_0(pr)}{\left[p^2 - (p_0^+)^2\right]\left[p^2-(p_o^-)^2\right]} = -\frac{i}{2\beta}\bigg\{ \xi^+K_0(-ip_0^+r) - \xi^-K_0(ip_0^-r)\bigg\}\\
I_G^2 &= \int_0^\infty dp \frac{p^2 J_1(pr)}{\left[p^2 - (p_0^+)^2\right]\left[p^2-(p_o^-)^2\right]} = -\frac{1}{2\beta}\left[p_0^+K_1(-ip_0^+r) +p_0^-K_1(ip_0^-r)\right],\\
\end{align}
where $K_i(x)$ is the modified Bessel function of the second kind, and $\xi^\pm = \frac{(p_0^\pm)^2}{2m} - \frac{p_F^2}{2m}$. Here $p_0^\pm$ are the zeroes of the denominator in the respective Green's function (indices $\lambda$, $\lambda^\prime$ have been suppressed). Specifically, we have for out-of-plane superconductivity
\begin{equation}
\begin{split}
p_0^\pm &= \sqrt{p_F^2 - 2m^2\Delta^2 \pm i2m\Delta\sqrt{p_F^2 - m^2\Delta^2 - \omega_\lambda^2/\Delta^2}}\\
\beta &= 2m\Delta\sqrt{p_F^2 - m^2\Delta^2 - \omega_\lambda^2/\Delta^2}
\end{split}
\end{equation} 
and for in-plane SC
\begin{equation}
\begin{split}
p_0^\pm &= \sqrt{p_F^2 - 2m^2\Delta^2 \pm i2m\Delta\sqrt{p_F^2 - m^2\Delta^2 - \omega_\lambda^2/\Delta^2}}\\
\beta &=  2 m \Delta \sqrt{p_F^2 - \left(m\Delta - \frac{\lambda\lambda^\prime B}{\Delta}\right)^2 - \frac{\omega^2-B^2}{\Delta^2}}.
\end{split}
\end{equation}
The result \eqref{app:eq:IG0} is obtained through residue integration upon representing the Bessel function as an integral; the others can consequently be deduced through recurrence relations. We have in this work assumed that the argument of the square root in $\beta$ is positive in all cases; to first order in $\Delta/v_F$, this is equivalent to requiring that the energies lie within the bulk gap.

The integrals in $W_j$ can be solved in a similar manner. We first consider the case $r = 0$. This can be easily solved by factorizing the denominator as above, yielding
\begin{align}
W^{\text{a}}_j(0) &= (-1)^{j}\sum_{\lambda = \pm 1} \mc P^\sigma_\lambda\lambda\tau_j \frac{\Delta m^2\left[p_0^+ + p_0^-\right]}{4B}\\
W^{\text{p}}_j(0) &= -\sum_{\lambda,\lambda^\prime = \pm 1} \mc P^\sigma_\lambda\mc P^\tau_{\lambda^\prime}\lambda\sigma_j\tau_x \frac{\Delta m^2\left[p_0^+ + p_0^-\right]}{4B}.
\end{align}
Consider then the case $r > 0$. The angular integral is most conveniently handled by pulling out a derivative with respect to $r_j$, upon which again we obtain a number of integrals over Bessel functions:
\begin{align}
W_j^\text{a}(\vec r) &=  \frac{i}{2\pi r}\sum_{\lambda = \pm 1}\mc P^\sigma_\lambda\left[\omega_\lambda r_j I_W^0 + \tau_z r_j I_W^1 - a_j I_W^0 - b_j I_W^3\right]\\
W_j^\text{p}(\vec r) &=  -\frac{i}{2\pi r}\sum_{\lambda,\lambda^\prime = \pm 1}\mc P^\sigma_\lambda\mc P^\tau_{\lambda^\prime}\bigg\lbrace\left[-\omega_{-\lambda}r_j + i\Delta\tau_x(\sigma_j - 2\frac{r_j}{r}\bs\sigma\cdot\vec{\hat r})\right]I_W^0 -\lambda^\prime r_j I_W^1 + i\Delta\tau_x \bs\sigma\cdot\vec{\hat r}r_j I_W^2\bigg\rbrace.
\end{align}
defined in terms of
\begin{align}
I_W^0 &= \int_0^\infty dp\frac{pJ_1(pr)}{\left[p^2 - (p_0^+)^2\right]\left[p^2-(p_o^-)^2\right]} =  \frac{\pi}{4\beta}\bigg[J_1(p_0^+r) + iH_1(p_0^+r) + J_1(p_0^-r) - iH_1(p_0^-r)\bigg]\\
I_W^1 &= \int_0^\infty dp\frac{p\xi_pJ_1(pr)}{\left[p^2 - (p_0^+)^2\right]\left[p^2-(p_o^-)^2\right]} = \frac{\pi}{4\beta}\bigg(\xi^+\left[J_1(p_0^+r) + iH_1(p_0^+r)\right] + \xi^-\left[J_1(p_0^-r) - iH_1(p_0^-r)\right] + \frac{2\beta}{\pi m}\bigg)\\
I_W^2 &= \int_0^\infty dp\frac{p^2J_0(pr)}{\left[p^2 - (p_0^+)^2\right]\left[p^2-(p_o^-)^2\right]} = \frac{\pi }{4\beta}\bigg(p_0^+\left[J_0(p_0^+r) + iH_0(p_0^+r)\right] + p_0^-\left[J_0(p_0^-r) - iH_0(p_0^-r)\right]\bigg)
\end{align}
where $H_i$ is the Struve function of the first kind, and the other parameters are the same as previously.

Finally we turn to the integrals in the previous appendix. In both cases, the angular integral is trivial and can be integrated out. Looking at the remaining integrals over the momentum, it is then immediately evident that
\begin{equation}
\begin{split}
I^1_{j} &= W_j(0)\\
I^2_{jk} &= \frac{\delta_{jk}}{2}G_0(0,\omega).
\end{split}
\end{equation}

\end{widetext}
\end{document}